\documentclass[submission,copyright,creativecommons]{eptcs}
\providecommand{\event}{ALP4IoT 2016} 
\usepackage{underscore}           

\usepackage{amsmath,amsthm}
\usepackage{graphicx}
\usepackage[usenames,dvipsnames,svgnames,table]{xcolor}
\usepackage{stmaryrd}
\usepackage{fancyvrb}
\usepackage{lipsum}

\newtheorem{defn}{Definition}

\newcommand{\FORGET}[1]{}

\newcommand{\PROGRAM}{\mathtt{P}}
\newcommand{\FUNCTION}{\mathtt{F}}

\newcommand{\e}{\mathtt{e}}

\newcommand{\fname}{\mathtt{d}}
\newcommand{\xname}{\mathtt{x}}

\newcommand{\bname}{\mathtt{b}}

\newcommand{\anyvalue}{\mathtt{v}}
\newcommand{\lvalue}{\ell}
\newcommand{\fvalue}{\phi}

\newcommand{\nbrK}{\mathtt{nbr}}
\newcommand{\repK}{\mathtt{rep}}
\newcommand{\ifK}{\mathtt{if}}

\newcommand{\letK}{\mathtt{let}\;}
\newcommand{\inK}{\;\mathtt{in}\;}

\newcommand{\toSymK}{\mathrm{\texttt{=>}}}

\newcommand{\deviceId}{\delta}

\newcommand{\dvalue}[0]{\mathrm{\Phi}}

\definecolor{dark-gray}{gray}{0}
\newcommand{\il}[1]{{\it \textcolor{dark-gray}{// #1}}} 
\newcommand{\km}[1]{{\textcolor{red}{#1}}} 
\newcommand{\fn}[1]{\textcolor{violet}{#1}} 
\newcommand{\pr}[1]{\texttt{\textcolor{blue}{#1}}} 

\newcommand{\ap}[1]{\langle #1 \rangle}
\newcommand{\bp}[1]{\left\lbrace #1 \right\rbrace}
\newcommand{\vp}[1]{\left\lvert #1 \right\rvert}

\DeclareMathOperator{\dist}{dist}

\newcommand{\MMM}{\mathcal{M}}

\title{Aggregate Graph Statistics}
\author{Giorgio Audrito
\institute{University of Torino, Italy}
\email{giorgio.audrito@unito.it}
\and
Ferruccio Damiani
\institute{University of Torino, Italy}
\email{ferruccio.damiani@unito.it}
\and
Mirko Viroli
\institute{University of Bologna, Italy}
\email{mirko.viroli@unibo.it}
}
\def\titlerunning{Aggregate Graph Statistics}
\def\authorrunning{G. Audrito, F. Damiani \& M. Viroli}

\begin{document}
\maketitle

\begin{abstract}
Collecting statistic from graph-based data is an increasingly studied topic in the data mining community. 
We argue that these statistics have great value as well in dynamic IoT contexts: they can support complex computational activities involving distributed coordination and provision of situation recognition.
We show that the HyperANF algorithm for calculating the neighbourhood function of vertices of a graph naturally allows for a fully distributed and asynchronous implementation, thanks to a mapping to the field calculus, a distribution model proposed for collective adaptive systems. This mapping gives evidence that the field calculus framework is well-suited to accommodate massively parallel computations over graphs. Furthermore, it provides a new ``self-stabilising'' building block which can be used in aggregate computing in several contexts, there including improved leader election or network vulnerabilities detection.
\end{abstract}

\section{Introduction}

Statistical informations are having an increasingly important role in handling massive graph-based data, such as the Web Graph or social network graphs. In this context, statistical summaries are able to detect local- or global-level features which can give approximate solutions to problems which would not be solvable exactly for graphs of that size. A particularly versatile statistic is the \emph{neighbourhood function} $N_G$, which calculates the number of vertices within a certain distance from a given vertex, and can be computed efficiently for massive graphs through the state-of-the-art HyperANF algorithm \cite{vigna:hyperanf}. Through $N_G$ several specific problems can be addressed \cite{palmer:anf}: among many others, ranking vertices by their \emph{harmonic centrality} (which is easily derivable from $N_G$).

Vertex ranking is particularly relevant for distributed IoT scenarios, and has already been considered in this context for fixpoint-based measures such as \emph{PageRank} \cite{lafuente:fixpoint}. These measures can help selecting ``first-class'' vertices (which could fruitfully be promoted to local communication hubs), or weaknesses in the network infrastructure (vertices with few close-range neighbours). However, algorithms suitable for fully-distributed IoT scenarios need to satisfy several non-trivial requirements, such as: resilience to network changes, continuous adaptation to input evolution (formalizable by \emph{self-stabilisation} \cite{VBDP-SASO2015}), near-constant time complexity for each computational unit, locality of interactions (obtainable with the \emph{aggregate programming} paradigm \cite{BPV-COMPUTER2015} for designing individual behaviour from a collective viewpoint). Porting algorithms to the IoT is thus a non-trivial task which may require extensive work where natural translations are not possible or sufficiently performing.

In this paper, we provide a natural translation of HyperANF in Field Calculus \cite{DVB-SCP2016,forte2015}, a tiny functional language for aggregate programs. On the one hand, this translation gives an example suggesting that field calculus is well-suited for expressing massively parallel computations over graphs. On the other hand, we obtain in this way a self-stabilising and local program which can be applied to solve problems such as improved leader election or network vulnerabilities detection. Section \ref{sec:background} introduces the concepts needed to express the translation, with an overview of the related works. Section \ref{sec:aggraphstat} provides the translation, together with a motivation outlining its possible use cases.

\section{Background and Related Work} \label{sec:background}

\subsection{Graph Statistics} \label{ssec:graphstat}

Many different techniques for collecting statistics from graph-based data are being studied in the data mining community, each able to address different questions. In this paper, we focus on statistics derivable from (variations of) the \emph{neighbourhood function} of a graph.

\begin{defn}[Neighbourhood Function]
	Let $G = \ap{V,E}$ be a graph with $n$ vertices and $m$ edges. The \emph{generalized individual neighbourhood function} $N_G(v, h, C)$, given $v \in V$, $h \geq 0$ and $C \subseteq V$, counts the number of vertices $u \in C$ which lie within distance $h$ from $v$. In formulas, $N_G(v, h, C) = \vp{ \bp{ u \in C : ~ \dist(v,u) \leq h } }$.
\end{defn}

Elaborations of the $N_G$ values have been used to answer many different questions \cite{vigna:hyperanf,palmer:anf}, including: graph similarity, vertex ranking, robustness monitoring, network classification. Since exact computation of $N_G$ is impractical, requiring $O(nm)$ time in linear memory and $O(n^{2.38})$ time in quadratic memory, effort has been spent in developing fast algorithms approximating $N_G$ up to a desired precision.

In Palmer et al. \cite{palmer:anf}, the problem of computing $N_G$ is reduced to that of succinctly maintaining size estimates for sets upon set unions. Define $\MMM_G(v, h, C)$ as the set of vertices in $C$ within distance $h$ from $v$, so that $N_G(v, h, C) = \vp{ \MMM_G(v, h, C) }$. Notice that $\MMM_G$ can be computed recursively as:
\[
\MMM_G(v, h, C) = \bigcup_{(vu) \in E} \MMM_G(u, h-1, C)
\]
where $\MMM_G(v, 0, C) = \bp{v} \cap C$. If we represent the sets $\MMM_G(v, h, C)$ through succinct counters, the previous formula translates into an algorithm computing estimations of $N_G$. Vigna et al. \cite{vigna:hyperanf} later improved the original algorithm by using a more effective class of estimators, the \emph{HyperLogLog counters} \cite{flajolet:hyperloglog}, by expressing the ``counter unions'' through a minimal number of broadword operations, and by engineering refined parallelisation strategies.
HyperLogLog counters maintain size estimates with asymptotic relative standard deviation $\sigma/\mu \leq 1.06 / \sqrt{k}$, where $k$ is a parameter, in $(1 + o(1)) \cdot k \cdot \log \log (n/k)$ bits of space. Updates are carried out through $k$ independent ``max'' operations on $\log \log (n/k)$-sized words. It follows that $N_G$, given a fixed precision, can be computed in $O(nh)$ time and $O(n \log \log n)$ memory, allowing it to be applied on very large graphs such as the Facebook graph \cite{vigna:facebook}.

\subsection{Field Calculus}

The Field Calculus is a tiny functional language for formally and practically expressing aggregate programs: a detailed account of it is given in \cite{DVB-SCP2016,forte2015}---we hereby recollect its most basic characteristics in order to be able to use it as common language to express the algorithms used in this paper with actual executable code. 


In field calculus a program  $\PROGRAM$ consists of a sequence of function declarations $\overline{\FUNCTION}$  and of a main expression $\e$---following~\cite{FJ}, the overbar notation denotes metavariables over sequences: e.g., $\overline\e$ ranges over sequences of expressions $\e_1, \ldots, \e_n$ with $n\ge 0$.
On each device $\deviceId$ the expression $\e$ evaluates to a value $\anyvalue$ that may depend on the state of $\deviceId$ (values sensed by $\deviceId$, the result of previous evaluation, and information coming from neighbours). Therefore the expression $\e$ induces a \emph{computational field} $\dvalue$, which can be represented as a time-varying map $\deviceId_1 \mapsto \anyvalue_1,\ldots, \deviceId_n \mapsto \anyvalue_n$, assigning a value $\anyvalue_i$ to each device $\deviceId_i$ in a network. 
Each device $\deviceId$ updates its value (by evaluating $\e$) in asynchronous computational rounds.

The syntax of an expression $\e$ comprises several constructs, which we briefly present. Firstly, $\e$ can be a variable $\xname$, used as function formal parameter, or a value $\anyvalue$ which in turn could be either a \emph{neighbouring field value} $\fvalue$ (associating to each device a map from neighbours to local values), or a \emph{local value} $\lvalue$ (built-in functions $\bname$, user-defined functions $\fname$, anonymous functions $(\overline\xname) \toSymK \e$, or built-in values).
Secondly, $\e$ can be a function call $\e(\overline\e)$ where an expression of functional type $\e$ is applied by-value to arguments $\e_1,\ldots\e_n$; or an $\ifK$-expression $\ifK ( \e_0 ) \{ \e_1 \} \{ \e_2 \}$ modelling domain restriction, which computes $\e_1$ in the devices where $\e_0$ is true, and $\e_2$ on the others.

Finally, $\e$ can be a $\repK$-expression or an $\nbrK$-expression, which are the two most characteristic constructs of field calculus: a $\repK$-expression $\e = \repK(\e_1)\{(\xname) \toSymK{} \e_2\}$ models time evolution and state preservation, by repeatedly applying the anonymous function $(\xname) \toSymK{} \e_2$ to the value computed for $\e$ in the previous round, starting from $\e_1$ if no previous value is available; an an $\nbrK$-expression $\nbrK\{\e\}$ models neighbourhood observation, by producing a neighbouring field value that represents an ``observation map'' of neighbour's values for expression $\e$, namely, associating to each device $\deviceId$  (that has evaluated $\nbrK\{\e\}$ during its last update) a map from neighbours to their latest evaluation of $\e$.

In order to conveniently present the code of the algorithms, we extend the core calculus with few additional constructs, which can be defined as syntactic sugar in terms of the presented ones: let-binding expressions $\letK \xname = \e_1 \inK \e_2$ (operationally equivalent to $((\xname) \toSymK \e_2)\e_1$), binary operators in infix notation, and the notation $[\e_1,\ldots,\e_n]$ for  tuples.

\section{Aggregate Graph Statistics} \label{sec:aggraphstat}


In the next subsection we shall show how the HyperANF algorithm presented in Section \ref{ssec:graphstat} can be naturally translated into field calculus. This transformation suggests that the field calculus is a natural framework for parallel computations, so that a future cloud-based implementation of it \cite{VCP-UBICOMPW2016} could be used to address a relevant class of ``traditional'' massively parallel computations.
However, this translation has further relevance in the aggregate computing context, since the resultant algorithm is able self-adjust to dynamic changes in inputs (i.e., has the \emph{self-stabilisation} property and belongs to the \emph{self-stabilising fragment} of the calculus \cite{VBDP-SASO2015}). In fact, HyperANF provides a new \emph{building block} which can be used to produce various dynamic vertex rankings, able to classify vertices by features of their neighbourhoods. This rankings can in turn be used either to recognise network vulnerabilities (vertices with few short-range neighbours and many long-range neighbours), or to elect leaders for other coordination mechanisms (vertices with high degrees of ``centrality'').
For example, $N_G$ can be used to compute the \emph{harmonic centrality} of each vertex through the formula $H(v) = \sum_{h>0} N_G(v, h, V)/h$. Vertices with high harmonic centrality are best-suited to be elected as leaders for coordination mechanisms, since they are connected to many other vertices through a small number of hops.

\subsection{HyperANF in Field Calculus}


Assume that we are given a \texttt{HyperLogLog} type, which can be constructed out of numerical elements, composed through a union operator, and inspected for size through function \texttt{estimate}. We can express HyperANF through the following simple field calculus code, computing a list of floating-point estimations $\ap{N_G(\deviceId, i, C): ~ i=0 \ldots h}$ in each device $\deviceId$, provided that \texttt{source} is true if and only if $\deviceId \in C$.
\begin{Verbatim}[fontsize=\fontsize{9pt}{10pt}, frame=single, commandchars=\\\{\}, codes={\catcode`$=3\catcode`^=7\catcode`_=8}]
\km{def} \fn{HyperANF}(h, source) \{ \il{ has type: (num, bool) $\rightarrow$ (HyperLogLog, list(float))}
  \km{if} (h == 0) \{
    \km{let} c = \km{if} (source) \{\fn{HyperLogLog}(\pr{uid})\} \{\fn{HyperLogLog}()\} \km{in} [c, \pr{list}(\fn{estimate}(c))]
  \} \{
    \km{let} r = \fn{HyperANF}(h-1, source) \km{in}
    \km{let} c = \pr{unionhood}( \km{nbr}\{\pr{1st}(r)\} ) \km{in} [c, \pr{cons}(\pr{2nd}(r), \fn{estimate}(c))]
\} \}
\end{Verbatim}

In Vigna et al. \cite{vigna:hyperanf}, non-trivial effort was made to allow for efficient parallel computation of HyperANF, since the algorithm did not fit inside existing parallel computation frameworks (e.g., MapReduce \cite{google:mapreduce}). Thus, it is worth noting that the field calculus is instead able to easily capture the algorithm, allowing for an ``automatic'' parallelisation and naturally extending the original algorithm to an ``on-line'' context (i.e., self-adapting to changes of either network structure or source vertices).

Function \texttt{HyperANF} can be used to implement real-time vertex ranking calculations for efficient leader election or network weaknesses detection. As a motivating example, we can integrate the leader election routine \texttt{S} \cite{BV-FOCAS2014} with HyperANF, by passing harmonic centrality instead of a random UID to the same inner function \texttt{BreakUsingUIDs}, thus favouring central devices to be elected as leaders.
\begin{Verbatim}[fontsize=\fontsize{9pt}{10pt}, frame=single, commandchars=\\\{\}, codes={\catcode`$=3\catcode`^=7\catcode`_=8}]
\km{def} \fn{HarmonicCentrality}(n) \{
  \km{if} (\pr{length}(n) <= 1) \{ 0 \} \{ \pr{head}(n)/(\pr{length}(n)-1) + \fn{HarmonicCentrality}(\pr{tail}(n)) \}        \}
\km{def} \fn{LeaderElection}(grain, metric, hmax, source) \{
  \fn{BreakUsingUIDs}(\fn{HarmonicCentrality}( \pr{2nd}(\fn{HyperANF}(hmax, source)) ), grain, metric)         \}
\end{Verbatim}

\bibliographystyle{eptcs}
\bibliography{long}
\end{document}